\documentclass[prl,twocolumn,superscriptaddress,showpacs,floatfix,long	bibliography]{revtex4-2}
\usepackage{graphicx}
\usepackage{bm}
\usepackage{color,soul}
\usepackage{amsmath,amssymb,amsfonts,float,graphics,epsfig,epstopdf,color,verbatim,tabularx,bm,multirow,appendix,hyperref}
\begin{document}
% Multiple \preprint coands are allowed.
%\preprint{}

\title{Nonlocal effects of low-energy excitations in quantum-spin-liquid candidate Cu$_3$Zn(OH)$_6$FBr}
\author{Yuan Wei}
\affiliation{Beijing National Laboratory for Condensed Matter Physics, Institute of Physics, Chinese Academy of Sciences, Beijing 100190, China}
\affiliation{School of Physical Sciences, University of Chinese Academy of Sciences, Beijing 100190, China}
\author{Xiaoyan Ma}
\affiliation{Beijing National Laboratory for Condensed Matter Physics, Institute of Physics, Chinese Academy of Sciences, Beijing 100190, China}
\affiliation{School of Physical Sciences, University of Chinese Academy of Sciences, Beijing 100190, China}
\author{Zili Feng}
\affiliation{Beijing National Laboratory for Condensed Matter Physics, Institute of Physics, Chinese Academy of Sciences, Beijing 100190, China}
\affiliation{Institute for Solid State Physics, University of Tokyo, Kashiwa 277-8581, Japan}
\author{Yongchao Zhang}
\affiliation{Beijing National Laboratory for Condensed Matter Physics, Institute of Physics, Chinese Academy of Sciences, Beijing 100190, China}
\affiliation{School of Physical Sciences, University of Chinese Academy of Sciences, Beijing 100190, China}
\author{Lu Zhang}
\affiliation{Beijing National Laboratory for Condensed Matter Physics, Institute of Physics, Chinese Academy of Sciences, Beijing 100190, China}
\affiliation{School of Physical Sciences, University of Chinese Academy of Sciences, Beijing 100190, China}
\author{Huaixin Yang}
\affiliation{Beijing National Laboratory for Condensed Matter Physics, Institute of Physics, Chinese Academy of Sciences, Beijing 100190, China}
\affiliation{School of Physical Sciences, University of Chinese Academy of Sciences, Beijing 100190, China}
\affiliation{Yangtze River Delta Physics Research Center Co., Ltd., Liyang, Jiangsu, 213300, People’s Republic of China}
\author{Yang Qi}
\affiliation{State Key Laboratory of Surface Physics, Department of Physics, Fudan University, Shanghai 200433, China}
\author{Zi Yang Meng}
\affiliation{Beijing National Laboratory for Condensed Matter Physics, Institute of Physics, Chinese Academy of Sciences, Beijing 100190, China}
\affiliation{Department of Physics and HKU-UCAS Joint Institute of Theoretical and Computational Physics, The University of Hong Kong, Pokfulam Road, Hong Kong, China}
\author{Yan-Cheng Wang}
\email{wangyc@cumt.edu.cn}
\affiliation{School of Materials Science and Physics, China University of Mining and Technology, Xuzhou 221116, China}
\author{Youguo Shi}
\email{ygshi@iphy.ac.cn}
\affiliation{Beijing National Laboratory for Condensed Matter Physics, Institute of Physics, Chinese Academy of Sciences, Beijing 100190, China}
\affiliation{Songshan Lake Materials Laboratory , Dongguan, Guangdong 523808, China}
\author{Shiliang Li}
\email{slli@iphy.ac.cn}
\affiliation{Beijing National Laboratory for Condensed Matter Physics, Institute of Physics, Chinese Academy of Sciences, Beijing 100190, China}
\affiliation{School of Physical Sciences, University of Chinese Academy of Sciences, Beijing 100190, China}
\affiliation{Songshan Lake Materials Laboratory , Dongguan, Guangdong 523808, China}
\begin{abstract}
We systematically study the low-temperature specific heats for the two-dimensional kagome antiferromagnet, Cu$_{3}$Zn(OH)$_6$FBr. The specific heat exhibits a $T^{1.7}$ dependence at low temperatures and a shoulder-like feature above it. We construct a microscopic lattice model of $Z_2$ quantum spin liquid and perform large-scale quantum Monte Carlo simulations to show that the above behaviors come from the contributions from gapped anyons and magnetic impurities. Surprisingly, we find the entropy associated with the shoulder decreases quickly with grain size $d$, although the system is paramagnetic to the lowest temperature. While this can be simply explained by a core-shell picture in that the contribution from the interior state disappears near the surface, the 5.9-nm shell width precludes any trivial explanations. Such a large length scale signifies the coherence length of the nonlocality of the quantum entangled excitations in quantum spin liquid candidate, similar to Pippard's coherence length in superconductors. Our approach therefore offers a new experimental probe of the intangible quantum state of matter with topological order.
\end{abstract}

\maketitle
The pursuit of quantum spin liquids (QSLs) has becoming one of the main themes in condensed matter physics and quantum material research, although by now few successes have been achieved~\cite{BroholmC20}. The central idea of a QSL lies in the fact that it is a quantum paramagnet with an anomalously high degree of entanglement, characterized by the nonlocal nature of the fractionalized excitations therein~\cite{SavaryL17,ZhouY17,WenXG17,WenXG19,BroholmC20}. Although of fundamental importance, these nonlocal effects are extremely hard to identify experimentally, this is because most of the experimental techniques only measure symmetry-breaking and local objects and thus can only probe QSL physics indirectly \cite{SavaryL17,BroholmC20}. Due to such indirectness, these experimental results in turn require precise theoretical modeling and calculations of frustrated quantum magnets to nail down the underlying physics, which is also very hard to achieve due to the lack of controlled quantum many-body computation methodologies for such strongly correlated systems. These difficulties force one to think of experimental probes that can directly measure the nonlocal excitations of a QSL material as the cases in the fractional quantum Hall effect \cite{MuEuenPL90,WangJK91}. Inspired by Pippard's coherence length for superconductors originated from the nonlocal response of Cooper pairs \cite{PippardAB53}, in this work, we have succeeded in probing a similar length scale existing for the nonlocal excitations of a QSL candidate material from nanoscale sample heat capacity measurements.

%One thus expect that a similar length scale may exist for the nonlocal excitations of a QSL.

The material Cu$_{3}$Zn(OH)$_6$FBr studied here belongs to the Cu$_{4-x}$Zn$_x$(OH)$_6$FBr system, which has the Cu$^{2+}$ ions with $S$ = 1/2 forming two-dimensional (2D) kagome layers. The $x$ in the molecular formula  denotes the content of Zn$^{2+}$ ions that substitute Cu$^{2+}$ ions between the kagome layers. The barlowite Cu$_4$(OH)$_6$FBr shows an antiferromagnetic (AFM) order at $\sim$ 15 K \cite{HanTH14,FengZL18,TustainK18}. The long-range AFM order disappears above $x \approx$ 0.4 while the short-range magnetic correlations from the interlayer spins can survive up to 0.82 \cite{FengZL18,WeiY20,TustainK20}. Experimental studies on Cu$_3$Zn(OH)$_6$FBr reveal typical fingerprints for a QSL in kagome antiferromagnet~\cite{JJWen2019}, including no long-range order down to 50 mK although its exchange interactions are about 200 K, the gap behavior of the NMR Knight shift, and spin continuum in neutron spectra which also suggest the existence of a gap \cite{FengZL17,WeiY17}. These results make Cu$_3$Zn(OH)$_6$FBr a promising candidate as a QSL with a $Z_2$ topological order~\cite{XGWen2017}. However, as in its siblings herbertsmithite ZnCu$_3$(OH)$_6$Cl$_2$ and Zn-doped claringbullite Cu$_3$Zn(OH)$_6$FCl \cite{NormanMR16,ShoresMP05,VriesMA08,HanTH12,HanTH16,FuM15,Kimchi2018,KhuntiaP20,ZLFengFCL2019,JJWen2019}, all the experimental results are obtained from local probes and the identification of the nonlocal excitations and entanglement therein are inferred indirectly. A direct observation of the nonlocal behaviors is critically in need to rule out alternative possibilities.

In this work, we studied the specific heat of Cu$_{3}$Zn(OD)$_6$FBr, which shows a shoulder at low temperatures followed by a quick drop with decreasing temperature. While this behavior resembles the gap behavior for its low-energy excitations, a power-law temperature dependence is found when $T$ goes to zero. We theoretically show that this behavior can be understood by the couplings between the visons and magnetic impurities. Moreover, the low-temperature specific heat from kagome planes shows a strong grain-size dependence, which leads to a large length scale of about 5.9 nm within a simple core-shell picture. This means that the low-energy excitations in Cu$_3$Zn(OH)$_6$FBr should have a coherence length due to their nonlocality, consistent with the existence of visons.

Polycrystalline Cu$_{3}$Zn(OD)$_6$FBr (Cu3Zn) were synthesized by the hydrothermal method as reported previously \cite{FengZL18}. Here deuterium was used for measuring the specific heats below 1.8 K due to its smaller nuclear Schottky anomaly. The samples were ground by hand in an agate mortar to obtain different grain sizes. The sizes of the grains were directly measured by the transmission  electron microscope (TEM). The samples were pressed into small pellets, whose specific heats were measured on a physical properties measurement system with the dilution refrigerator (Quantum Design, 9T).

We also constructed a microscopic model on the kagome lattice which is well-known for having a $Z_2$ quantum spin liquid ground state~\cite{BalentsL02,IsakovSV11,IsakovSV12,WangYC17,YCWang2017,GYSun2018,YCWang2020}. The kagome $Z_2$ state in this model acquires both spinon and vison anyonic excitations. We add impurities to the model and couple the impurity to the kagome plane which is inside the quantum spin liquid ground state, the coupling is strong in the $S^z$ channel which has been proven to probe the vison-pair excitations~\cite{GYSun2018,YCWang2020}, i.e., the coupling connects the impurities with the vison excitations of the quantum spin liquid. We then compute the specific heat of the coupled system with exact unbiased quantum Monte Carlo (QMC) simulations~\cite{WangYC17,GYSun2018,YCWang2020}, both with and without external field, to explain the experimental findings.

\begin{figure}[tbp]
\includegraphics[width=\columnwidth]{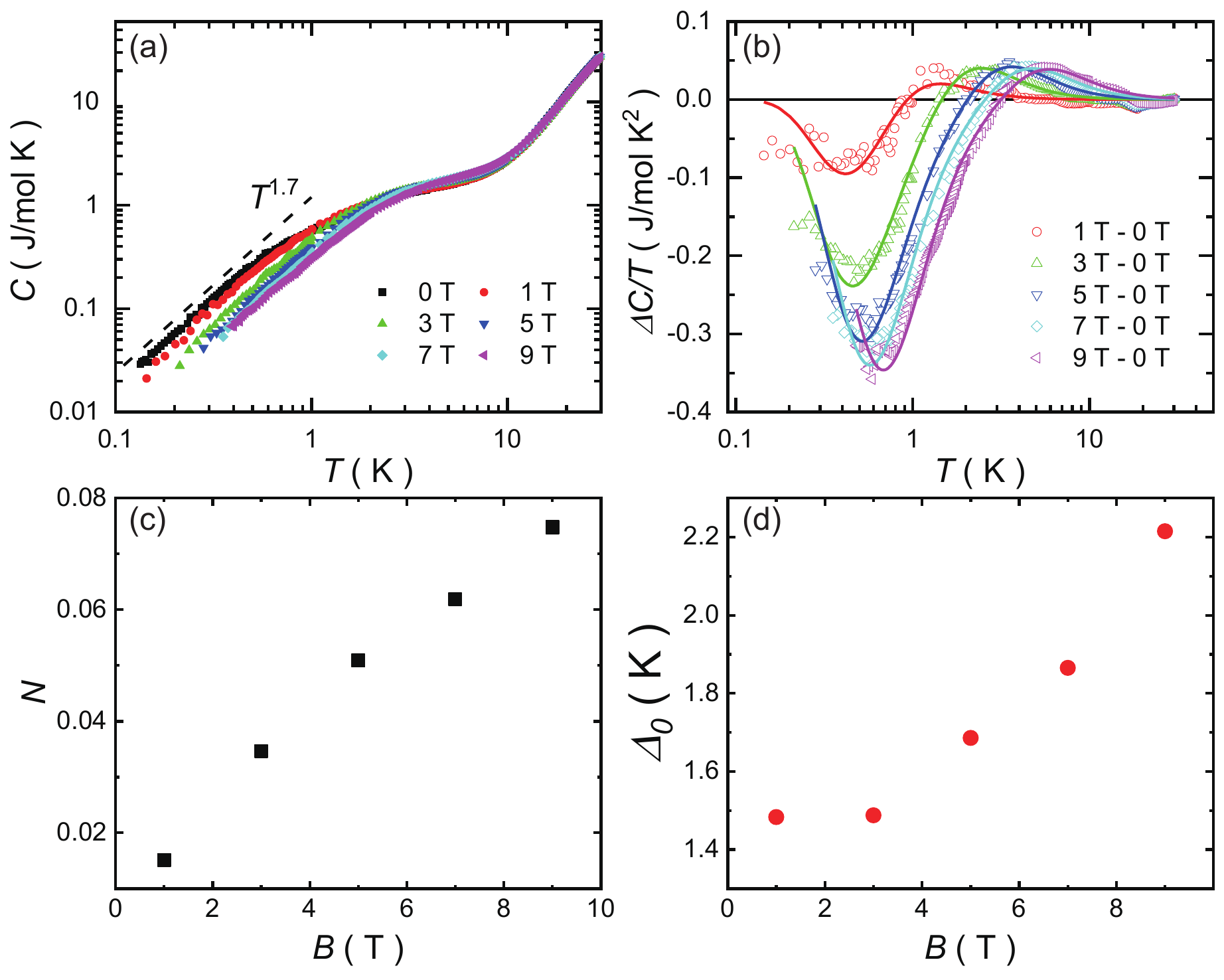}
 \caption{(a) The specific heat of bulk Cu3Zn as a function of the temperature at different fields. The dashed line shows the $T^{1.7}$ temperature dependence. (b) The temperature dependence of $\Delta C/T$. The solid lines are fitted results by the Schottky-anomaly function as discussed in the main text. (c) and (d) The field dependence of $N$ and $\Delta_0$ in the Schottky-anomaly function, respectively.}
\label{SCexp}
\end{figure}

Figure \ref{SCexp}(a) shows the temperature dependence of the specific heat of the as-grown Cu3Zn sample, labeled as "bulk", which has typical grain sizes of several micrometers. With decreasing temperatures, the specific heat exhibits a shoulder around 4 K. The entropy released below 6 K is about 2.5 J/mol K, which is larger than 0.4Rln2 and should mainly come from the low-energy magnetic excitations of the kagome lattice as the phonon contribution here becomes negligible \cite{VriesMA08,SchnackJ18}. As previous studies have suggested the gapped magnetic ground state for Cu3Zn \cite{FengZL17,WeiY17}, one expects that the gap behavior should also be observed in the specific heat. Instead, as shown in Fig. \ref{SCexp}(a), a power-law temperature dependence of the specific heat is found, i.e. $C \propto T^{\alpha}$, with the exponent $\alpha \sim$ 1.7 which decreases slightly with increasing field. These results are very similar with those in herbertsmithite \cite{VriesMA08} and indicate that the magnetic excitations are gapless.

The seemingly contradicting conclusions for the gapped nature of the magnetic ground states have also been found in the studies on herbertsmithite \cite{NormanMR16,ShoresMP05,VriesMA08,HanTH12,HanTH16,FuM15,Kimchi2018,KhuntiaP20}. There are proposals that the kagome antiferromagnet could be a gapless Dirac spin liquid \cite{RanY07}, but the power-law of specific heat we observed does not fit into this picture since both the exponent and the field dependence differ from the expectation of the Dirac spin liquid. From the gapped picture, the low-temperature specific heat has been explained as the simple sum of the gapped excitations from the kagome planes and the contribution from inter-kagome Cu$^{2+}$ ions as magnetic impurities \cite{NormanMR16,HanTH16}. The essence in this scenario is that the magnetic impurities are only weakly correlated with each other and have no interactions with the kagome spins. This means that the contribution from the magnetic impurities should be easily pushed to high temperatures with the application of a moderate magnetic field and a gapped behavior should be then revealed, which was however not observed in the experiment \cite{VriesMA08}.

Here we take the viewpoint that, although kagome spin system of Cu3Zn is gapped, the interactions between the intra- and inter-kagome spins have to be taken into account to understand the low-temperature specific heat. First, we illustrate that it is not correct to view the magnetic impurities as only weakly coupled. As suggested previously \cite{VriesMA08}, the specific heat of weakly coupled magnetic impurities can be derived by fitting the difference of the total specific heat C between field $B$ and zero field by the Schottky function, i.e., $\Delta C$ = $C_{SH}(B) - C_{SH}(0)$, where $C_{SH}(B)$ = $NR(\Delta/k_BT)^2\exp(\Delta/k_BT)/(1+\exp(\Delta/k_BT))^2$ with $\Delta$ = $\Delta_0 + g\mu_BB$ for spin 1/2. Here $N$ and $\Delta_0$ are the number of magnetic impurities and intrinsic energy level at zero field. Figure \ref{SCexp}(b) shows the fitting results for $\Delta C/T$, which suggests that the Schottky function indeed captures the essence of the field dependence of the specific heat. However, both $N$ and $\Delta_0$ show strong field dependence (Fig. \ref{SCexp}(c) \& \ref{SCexp}(d)), which is odd especially for $N$ as it has been used to determine the impurity concentration \cite{VriesMA08,WeiY17}. 

\begin{figure}[tbp]
\includegraphics[width=\columnwidth]{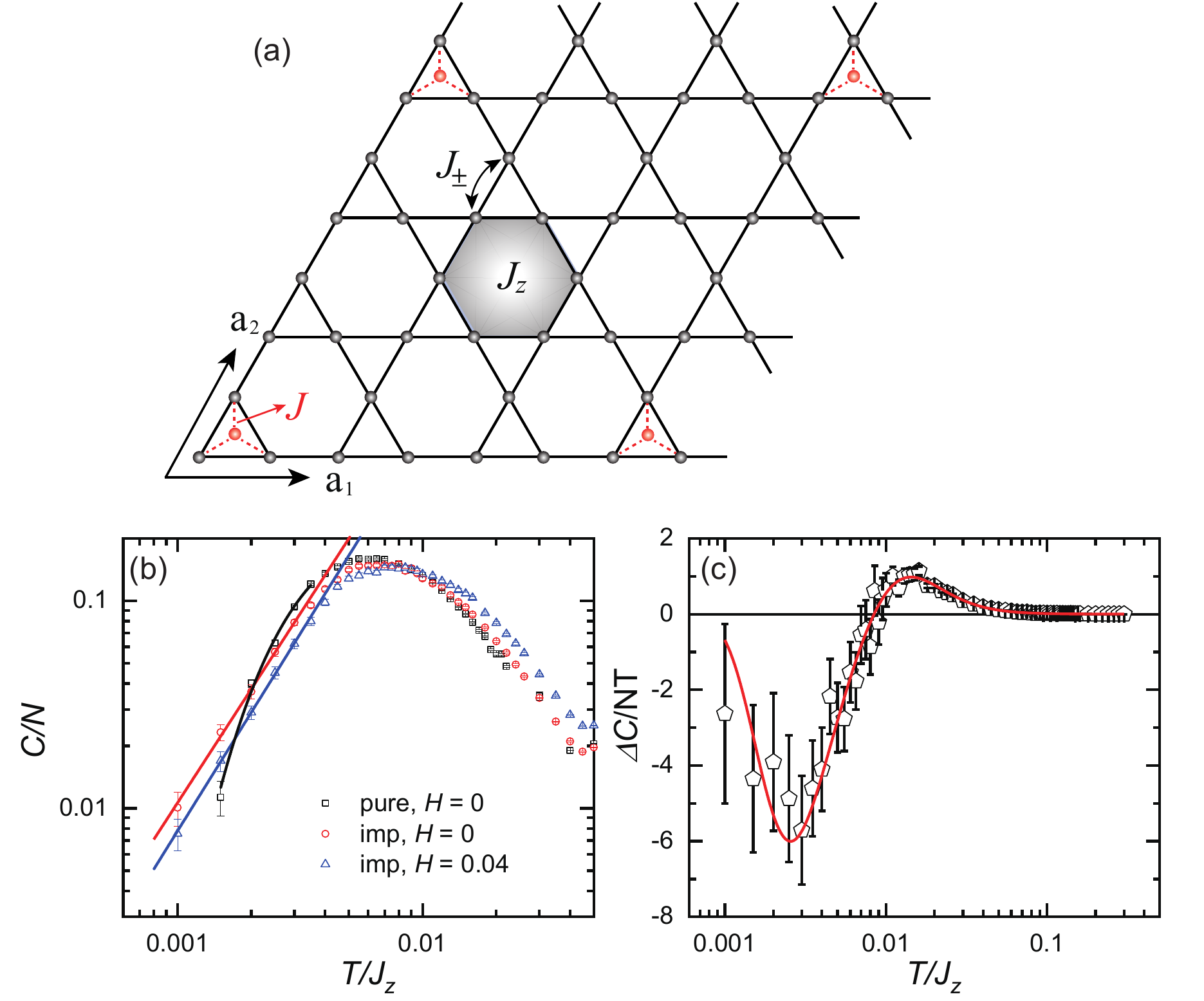}
 \caption{(a) Schematic diagram for the BFG model with 1/9 magnetic impurities (red dots). (b) Calculated low-temperature specific heat for the BFG model without (pure) and with (imp) impurities. The straight lines are fitted by the power function. The solid line for the pure sample is fitted by the gap function. (c) The difference of $\Delta C/T$ between field $H$ and zero. The solid line is fitted by the Schottky anomaly function.}
\label{SCtheory}
\end{figure}

The above results suggest that magnetic impurities cannot be simply treated as weakly correlated. We will show that the above results can be understood as from the gapped $Z_2$ QSL and the interplays between the impurities and the kagome spins. To this end, we constructed a microscopic kagome lattice model which is known to host $Z_2$ QSL ground state~\cite{BalentsL02,IsakovSV06,IsakovSV11,IsakovSV12}. The model is described in Fig. \ref{SCtheory}(a). The two types of anyonic excitations in the system are spinon and vison, with a gap of scale of $J_z\sim 1$ and $J^2_xy/J_z \sim 0.01$, as shown in our previous work~\cite{GYSun2018}.  To mimic the impurities, we further introduce spin-1/2 at the center of the triangles of the kagome lattice with 1/9 concentration, as shown in Fig. \ref{SCtheory}(a). 
We couple the impurities to the kagome plane as follows, $H_{imp} = J\sum_{\langle i,j\rangle'}\mathbf{S}_i \cdot \mathbf{S}_j + H \sum_i S_i^{z}$, where $J$ is the strength of antiferromagnetic heisenberg interaction,  $\langle \; \rangle'$ means the interaction between the impurity spins and the spins in a triangle of the kagome plane and $H$ is the longitudinally magnetic field for the total spins.

Without the impurity, the low-temperature specific heat of the $Z_2$ QSL model can be well fitted by a gap function $C \propto  T^{-1}\exp(-\Delta/T)$, as shown in Fig. \ref{SCtheory}(b). With the impurities, the gap behavior is replaced with the power law $T^{\alpha}$ with $\alpha$ changing from 1.82 at zero field to 1.9 at $H=0.04J_z$. We found that at $J=0.05J_z$, the obtained $\alpha$ is the closest to the experimental value. Increasing $J$ will increase $\alpha$, e.g., $\alpha \approx$ 2 for $J=0.1J_{z}$. The difference of $C/T$ between field $H$ and zero field is shown in Fig. \ref{SCtheory}(c) and can be well fitted by the Schottky anomaly function.

We note that for the $Z_2$ QSL in our model study, the specific heat of spinons shows a large hump at high temperature and is not shown here. The hump at low temperature comes from the visons. In Cu3Zn, the spinon and vison gaps may be much closer \cite{FengZL17,WeiY17}, so both of them should contribute to the low-temperature specific heat. In fact, our model simulation of the impurity on $Z_2$ QSL provides a qualitative understanding of the mechanism for the same coupling of impurities to the QSL in the actual material, which nicely capture the key features of the experimental low-temperature specific heat. It suggests that the interplay between the gapped topological anyons and magnetic impurities is crucial for the power-law and field-dependence behaviors of the specific heat.

\begin{figure}[tbp]
\includegraphics[width=\columnwidth]{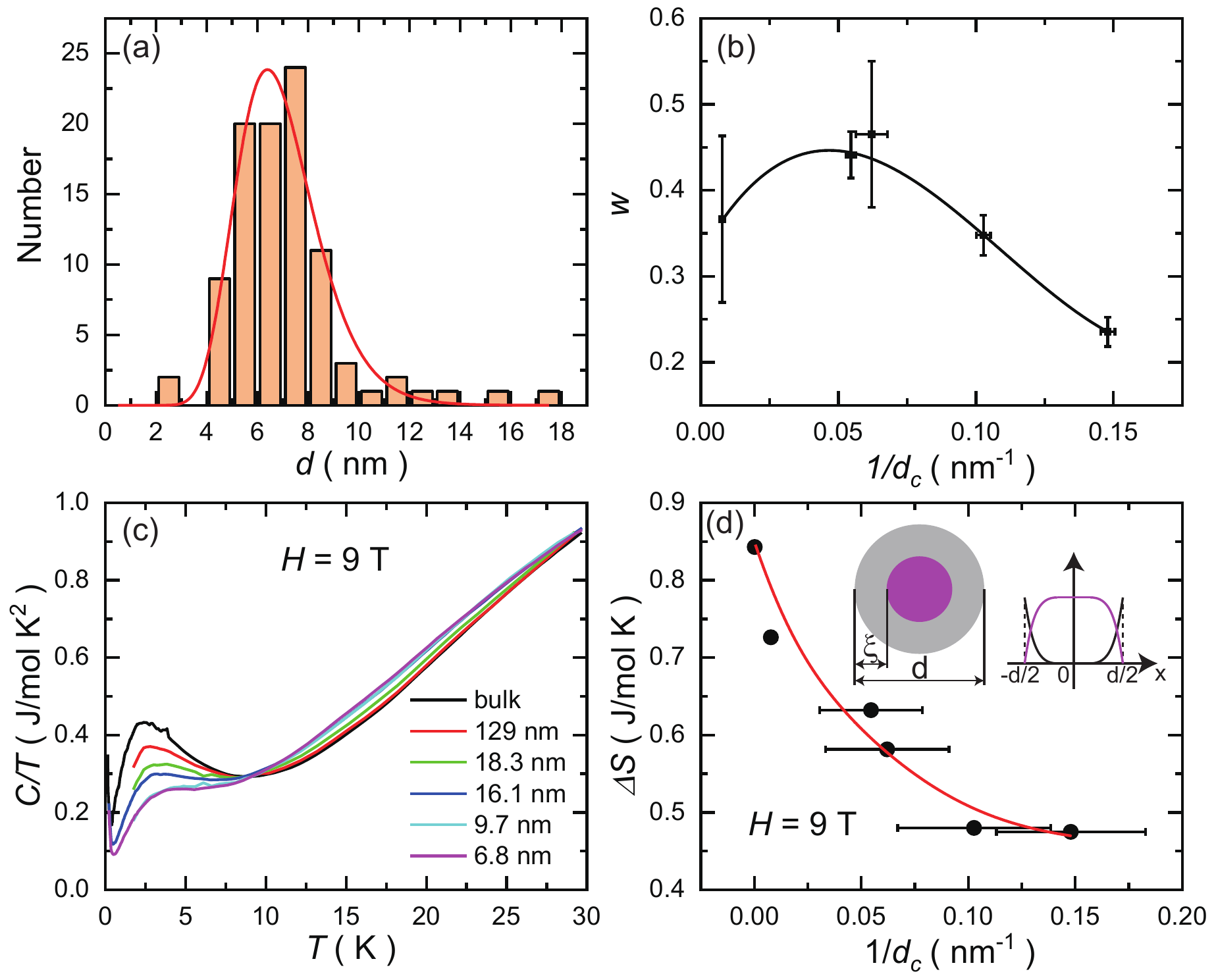}
 \caption{(a) Size distribution for a selected ground sample with the smallest average size. The solid line is fitted by the lognormal function. (b) The relationship between $w$ and $1/d_c$. The errors are from the lognormal fitting. The solid line is the fitting result of a polynomial function, which is only used for calculating the parameter $w$ in Eq. \ref{SCall} for the purpose of getting a smooth curve. (c) The $C/T$ at 9 T as a function of temperature for different sample. The values of the labels are $d_c$. (d) The $1/d_c$ dependence of the entropy change from 2 to 4 K. The horizontal bars are calculated from the log standard deviation $w$ of the lognormal function. The left inset shows a sketch of a two-dimensional circular sample with the diameter of $d$, whose surface area with the thickness of $\xi$ does not contribute to the low-temperature specific heat. The right inset illustrates that the surface state (black lines) exponentially decays into the sample's interior and in the meantime, the interior state (purple lines) quickly drops to zero near the surface.}
\label{grind}
\end{figure}

The above analysis, of course, only shows that the gapped anyons can explain the low-temperature specific heat, not vice versa. To truly reveal the unambiguous topological nature of the low-energy excitations from the gapped QSL, we further study the size dependence of the specific heat in Cu3Zn. After ground, the samples are thin plates with the thickness along the c-axis. While the thickness cannot be identified, its value is not concerned here since the spin system is 2D. With high-quality TEM images, we are able to determine the size $d$ of the plates to obtain the size distribution. Figure \ref{grind}(a) shows the result for the samples with the smallest average size, which can be fitted by the lognormal distribution function, $P(d) = Ae^{-(ln\frac{d}{d_c})^2/2w^2}/\sqrt{2\pi}wd$, where $d_c$ and $w$ are the mean value of the size and log standard deviation, respectively. We apply this analysis to all the samples and show their $d_c$ and $w$ in Fig. \ref{grind}(b). 

The low-temperature specific heat of all samples at 9 T are shown in Fig. \ref{grind}(c). Here the 9-T magnetic field is used to minimize the effects from magnetic impurities and really weakly correlated spins (such as orphan spins and surface spins) as much as possible, whose contributions are mostly pushed to high temperatures by the field. Figure \ref{grind}(d) shows the $1/d_c$ dependence of the entropy change $\Delta S$ from 2 to 4 K, where the major contribution is from the low-energy magnetic excitations from the kagome spin system. It is clear that $\Delta S$ strongly depends on the size of grains. 

This behavior can be simply understood as the core-shell effect, which assumes that the shell has different properties from the core, as shown in the inset of Fig. \ref{grind}(d), and thus does not contribute to the low-temperature specific heat anymore. Supposing that the surface state exponentially decays into the core as $\sim e^{-x/\xi}$, the contribution from the core is proportional to $1-e^{-x/\xi}$. Integrating the latter in two dimensions gives the fraction of the specific heat of the core as $f(d) = (1-\frac{2\xi}{d}(1-e^{-d/2\xi}))^2$. To account for the size distribution, we can further write down the following function,
\begin{equation}
C = (C_{bulk}-y_0)\frac{\int_0^\infty P(x)f(x)x^2dx}{\int_0^\infty P(x)x^2dx}+y_0,
\label{SCall}
\end{equation}
\noindent where $C_{bulk}$ is the specific heat for the as-grown sample and $y_0$ is introduced for the specific heat that has no size dependent. Since $w$ in the $P(x)$ is calculated accounting to the polynominal fitting in Fig. \ref{grind}(b), the fitting parameters in Eq. \ref{SCall} are just $\xi$ and $y_0$. The sold line in Fig. \ref{grind}(d) gives the fitting result, which gives $\xi \approx$ 5.9 nm. This length scale corresponds to about 9 in-plane lattice constant, or 18 nearest Cu-Cu distance. We note that the fitting can be further improved with more carefully considering the size distribution. Moreover, effects such as local lattice distortions may further affect the determination of $\xi$. However, the key result here is that the value of $\xi$ is still much larger than the Cu-Cu distance. 

The core-shell model has been widely applied in studying the ordered magnetism in nanoparticles \cite{KodamaRH97,MandalS09}, where the major idea is that the effect of uncompensated surface spins may extend into the core state and even cause fundamental change of ordered spin systems. However, for a trivial paramagnetic state, the effect of surface spins should be limited within a very short distance \cite{supp}. Therefore, the large size dependence of the low-temperature specific heat in Cu3Zn suggests that the low-energy excitations are strongly correlated as if there is some kind of order, which must have a nontrivial origin.

We recall that in superconductors, a similar core-shell picture applies for the Pippard's coherence length \cite{PippardAB53}. The underlying physics is that there is a minimum length over which a given change of the superfluid density of Cooper pairs can be made. This means that a layer with the thickness of the coherence length exists at the surface of a superconductor for the crossover from the superconducting state to the normal state or vacuum. The coherence length is in the order of $\upsilon/\Delta$, where $\upsilon$ and $\Delta$ are the Fermi velocity and superconducting gap, respectively. This comes from the nonlocal effects of Cooper pairs and thus the coherence length defines the intrinsic nonlocality of the superconducting state.

We argue that the length scale $\xi$ obtained in Cu3Zn should also come from the nonlocality of the low-energy excitations in the QSL state. Similar to superconductors, the emergent nonlocal excitations in QSLs may also have coherence lengths that define their nonlocality. It has been pointed out that real-life $s$-wave superconductors have a $Z_2$-topological order \cite{WenXG17,HanssonTH04,MorozS17}, and mean-field theories of $Z_2$ QSLs describe them as superconducting states of fermionic spinons, and the counterpart of Pippard's coherence length is the span of spin-singlet pairs in the resonant-valence-bond (RVB) picture~\cite{Anderson1973}. As suggested by previous results \cite{FengZL17,WeiY17}, Cu3Zn is a strong candidate for a $Z_2$ QSL or topological order. Considering that $J \sim$ 10 meV, it may give rise to a spin velocity in the order of 10 meV\AA and thus a gap in the order of 0.1 meV for a coherence length of 100 \AA. This gap value is consistent with the value estimated from the low-temperature shoulder of the specific heat. While the above arguments are reasonable for $Z_2$ QSLs, it is hard to apply them to U(1) QSLs with gapless excitation, since in mean-field theories, U(1) QSLs are described as metals, instead of superconductors, of spinons, and therefore do not have a coherence length.We also note that there is an alternative explanation to the large shell region, i.e., the shell region may be in a valence-bond solid phase with a much larger gap. However, this description also requires nonlocality of the system. 

In the end, we give a brief discussion on the nature of the low-temperature specific-heat shoulder in Cu3Zn. Numerical simulation revealed that the shoulder in kagome antiferromagnet mainly comes from singlet-singlet excitations \cite{SchnackJ18}. These $S$ = 0 excitations could be either trivial or non-trivial and the latter are visons \cite{HaegemanJ15}, which should show nonlocal effects because of their nonlocal topological nature. We note that $y_0$ in the above fitting is about 0.45 J/mol K, which is too large to be explained as phonons and may come from the trivial sector of the $S$ = 0 excitations. Nevertheless, the strongly size-dependent part of the specific heat may come from vison excitations, since we expect this contribution to vanish in the shell region because vison excitations, which can be viewed as vortices if the $Z_2$ QSL is viewed as a superconductor, cannot exist in the shell region within Pippard's coherence length.

Overall, our results put strong constraints on the nature of the ground state for Cu3Zn. We believe that the grain size thermodynamical measurements and analysis in this work open the avenue for direct measurements of the quantum entanglement in topological ordered states of matters and the concept of coherence length may provide a bulk property to address the ground-sate nature of the topological orders, as in superconductors.

\acknowledgments

S. L. thanks Prof. Yi Zhou, Prof. Guodong Li and Prof. Zheng Li for helpful discussions. This work is supported by the National Key Research and Development Program of China (Grants No. 2017YFA0302900, No. 2016YFA0300500,2020YFA0406003), the National Natural Science Foundation of China (Grants No. 11874401, No.11674406, No. 11961160699, No.11774399, No.11804383), the Strategic Priority Research Program(B) of the Chinese Academy of Sciences (Grants No. XDB33000000 and No. XDB28000000, No. XDB25000000, No. XDB07020000), the K. C. Wong Education Foundation (Grants No. GJTD-2020-01, No. GJTD-2018-01), Beijing Natural Science Foundation (No. Z180008).

\end{document}